\documentclass[12pt]{iopart}
\usepackage{amssymb,amscd,epsfig,bbold,stmaryrd,bm,dsfont}
\usepackage{graphicx}
\usepackage{color}
\usepackage{psfrag}
\usepackage{iopams}
\usepackage{dsfont}
\usepackage{setstack}
\usepackage{bm}

\newcommand{\Qint}{Q_\mathrm{int}}
\newcommand{\Qqp}{Q_\mathrm{qp}}

\begin{document}

\title{Measurement and dephasing of a flux qubit due to heat currents}

\author{Samuele Spilla$^{1,2,3}$, Fabian Hassler$^{2,4}$, and  Janine Splettstoesser$^{1,2}$}

\address{$^1$  Institut f\"ur Theorie der Statistischen Physik, RWTH Aachen University, D-52056 Aachen, Germany}
\address{$^2$ JARA-Fundamentals of Future Information Technology}
\address{$^3$ Dipartimento di Fisica e Chimica, Universit\`a di Palermo, I-90123 Palermo, Italy}
\address{$^4$ Institute for Quantum Information, RWTH Aachen University, D-52056 Aachen, Germany}

\begin{abstract}
We study a flux qubit, made of a superconducting loop interrupted by three
Josephson junctions, which is subject to a temperature gradient. We show that
the heat current induced by the temperature gradient, being sensitive to the
superconducting phase differences at the junctions, depends significantly on
the state of the qubit. We furthermore investigate the impact of the heat
current on the coherence properties of the qubit state.  We have found that
even small temperature gradients can lead to dephasing times of the order of
microseconds for the Delft-qubit design.
\end{abstract}

\maketitle

\section{Introduction}

The charge current through a superconducting weak link is sensitive to the
phase difference of the superconductor order parameters on either side of the
link.  In the absence of a bias voltage, a dissipationless Josephson current
flows through the link which is proportional to the sine of the phase
difference.  The origin of this supercurrent can be traced back to Andreev
reflection of incoming electrons and as such it is an interplay of
quasiparticles at the interface and the superconducting condensate on both
sides.

In 1965 Maki and Griffin~\cite{Maki65} theoretically predicted that also the
heat current flowing through a temperature-biased Josephson tunnel junction
is a periodic function of the phase difference between the electrodes.
Due to the invariance of the heat current under time reversal, it has an even parity with respect to the phase difference. The phase dependence of the heat current --- carried by
quasiparticles residing at energies outside of the energy gap of the
superconductor --- comes again from an interplay between these quasiparticles
and the superconducting condensate.

This effect has recently been demonstrated
experimentally~\cite{Giazotto12,Giazotto12a,Martinez13,Martinez13a,Giazotto13}.
A superconducting ring, namely a \emph{dc}-SQUID (superconducting quantum
interference device) with two Josephson junctions was exposed to a
temperature gradient.  The measurement of the resulting heat current as a
function of the magnetic flux penetrating the SQUID demonstrated the
sensitivity of the heat current to the phase differences across the
junctions.  In this way, the SQUID is operated as a heat modulator.

Heat transport through weak links in superconductors was theoretically studied
in great details~\cite{Zhao04,Giazotto06,Golubev13}, see also~\cite{Martinez13b} for a review on interference in heat transport and thermoelectric effects in superconducting weak links.  It has been found that
the heat current can be modulated by the applied phase gradient~\cite{Zhao03}.  Recent experiments have shown that weak links in
superconductors can be used to refrigerate small islands~\cite{Timofeev09} and trap hot quasiparticles~\cite{Nguyen13}.

An altogether different application of the phase sensitivity of the
supercurrent in superconducting rings is the realisation of a
persistent-current flux qubit where the phase sensitivity of the device is
used to implement qubit operations.  In particular, the Delft design of the
flux qubit consists in a superconducting loop interrupted by three Josephson
junctions.  It is furthermore characterised by the fact that the Josephson
coupling of one of the junctions is smaller by a factor $\alpha\simeq 0.75$~\cite{Mooij99}, which in actual implementations is made tuneable by replacing
this third junction by a split Josephson junction~\cite{Makhlin01}.  Another
important tuning parameter is the external flux $\Phi$ threading the loop.  If
the flux is close to half a superconducting flux quantum, $\Phi=h/4e$, the
superconducting system emulates a particle in a (shallow) double-well
potential, where the state in either well corresponds to a circulating
persistent current, either flowing clockwise or counterclockwise around the
loop.  These two states represent the qubit states of the device. 

In what follows, we will combine these two intriguing studies on the
phase-sensitivity in superconducting rings.  We are in particular interested
in the dependence of the heat current on the state of the persistent
current qubit.  We therefore investigate a superconducting ring with three
Josephson junctions subject to a temperature gradient.  We use a microscopic
description of the Josephson junctions in order to investigate the
phase-dependent heat current through them.  We will show that indeed the
heat current in a temperature-biased Delft qubit is sensitive to the qubit
state, with typical sensitivities of 4\%.

Beyond this, the state-sensitive heat current has an impact on the qubit
state.  With the help of a master equation approach, we investigate how the
temperature \textit{gradient} influences the dynamics of the qubit system.  We
determine the rate of coherence suppression which is shown to be given by the
rate with which the difference in heat currents at the two qubit states
accumulates an energy difference of approximately the gap energy.  The
difference in heat currents due to a thermal gradient depending on the qubit
state is hence demonstrated to be a qubit-state measurement.  Depending on the
temperature gradient, the associated typical dephasing times range from nano-
to microseconds with the proviso that the qubit is detuned from the ``sweet
spot'' of half a flux quantum, $\Phi=\Phi_0/2$, threading the superconducting
loop, to a typical "operation point'' of $\Phi=0.495\Phi_0$. This adds an additional contribution to the dephasing, which in general is attributed to non-equilibrium quasiparticles~\cite{Martinis09,Catelani11,Leppakangas12}. In the appendix,
we give a self-contained and detailed derivation of the results of Maki and
Griffin for the phase-sensitive heat current in a superconducting weak link.

\section{Model}

\subsection{Persistent current qubit (Delft qubit)}

\begin{figure}[t]
\begin{center}
\includegraphics[width=13.cm]{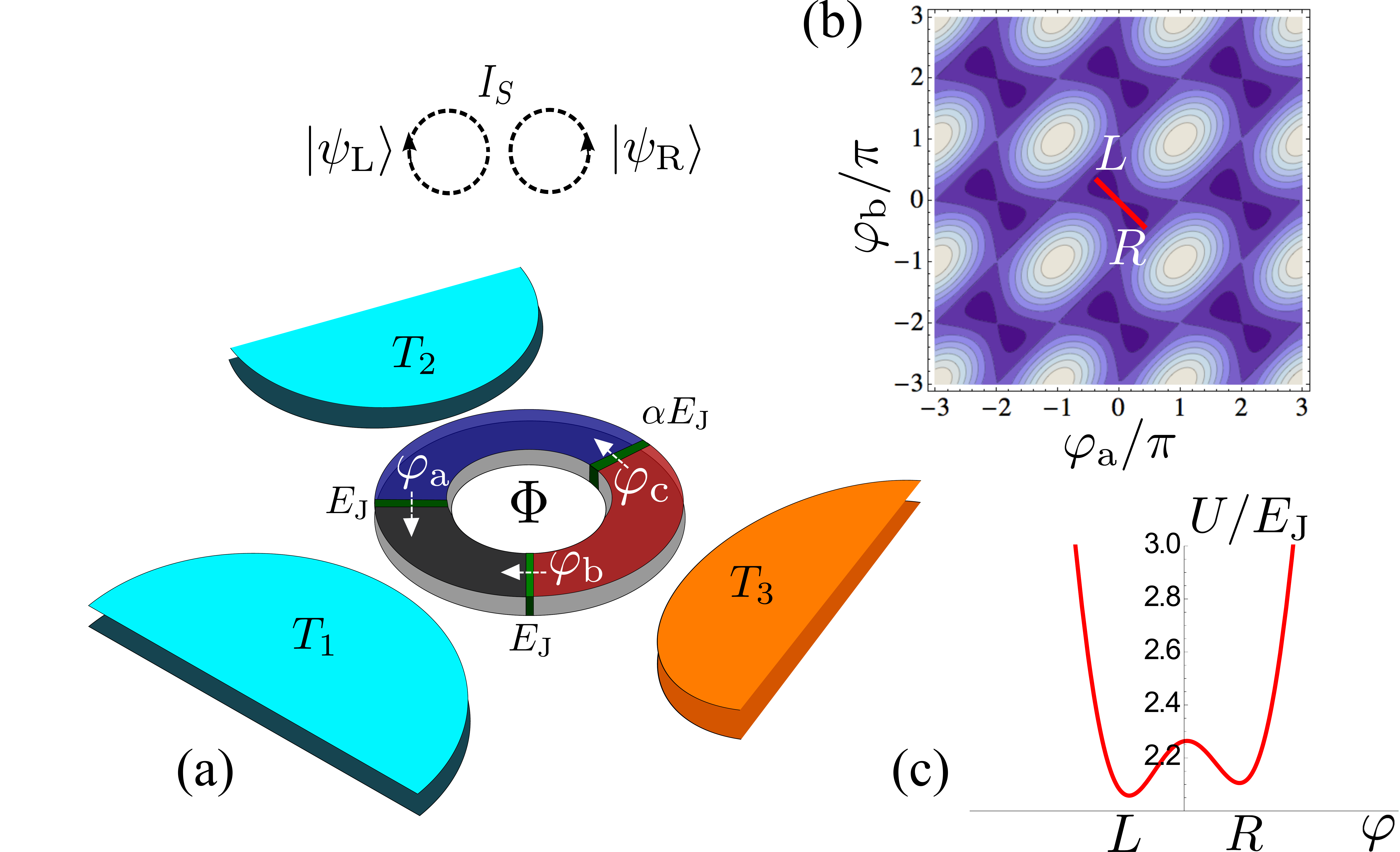}
\end{center}\caption{\textbf{(a)} Sketch of a persistent current qubit realised by a SQUID with three superconducting links, characterised in general by different phase differences, $\varphi_\mathrm{a},\varphi_\mathrm{b},\varphi_\mathrm{c}$. The  SQUID is penetrated by a magnetic flux $\Phi$. The direction of the  supercurrent circulating in the SQUID characterises the state of the persistent current qubit. The different sections of the SQUID are coupled to thermal baths with temperatures $T_1,T_2$ and $T_3$. \textbf{(b)} Potential landscape of the SQUID as a function of the phase differences for $\alpha=0.75$ and $\Phi/\Phi_0=0.495$. The potential is $2\pi$-periodic in $\varphi_\mathrm{a}$ and $\varphi_\mathrm{b}$ and it can be divided into equal square cells of side length $2\pi$. \textbf{(c)} Cut through the potential in the central cell of figure (b) along the line $\varphi=\varphi_\mathrm{a}=-\varphi_\mathrm{b}$ visualising the two minima corresponding to the qubit states of oppositely circulating currents.
\label{fig_DelftQubit}
}
\end{figure}

We investigate a persistent current qubit as it is sketched in
figure~\ref{fig_DelftQubit}. Such a qubit consists of  a superconducting loop
with three Josephson junctions, which encloses a flux
$\Phi$ supplied by an external magnetic field. 
The three junctions, $i=\mathrm{a,b,c}$, are in general characterised  by different Josephson energies $E_\mathrm{J}^{i}$, with $E_\mathrm{J}^{i} = I^{i}_\mathrm{crit}\Phi_0/2\pi$, where  $I^{i}_\mathrm{crit}$ is the critical current of the
junction and $\Phi_0=h/2e$ the superconducting flux quantum.  
Following the Delft-qubit design, we choose two of the junctions to be equivalent, i.e. having the same Josephson coupling energy
$E_\mathrm{J}^{\mathrm{a}}=E_\mathrm{J}^{\mathrm{b}}\equiv E_\mathrm{J}$, and the
third junction with a smaller Josephson energy
$E_\mathrm{J}^{\mathrm{c}}=\alpha E_\mathrm{J}$, where we have introduced the
asymmetry parameter $\alpha \leq 1$.  
The different phase differences, $\varphi_i$, across the junctions (the arrows
in figure~\ref{fig_DelftQubit} define the direction for a positive phase difference $\varphi_i$) are related to each other due to the fluxoid quantization around the superconducting  loop containing the junctions,\footnote{We here neglect loop inductances.}
\begin{equation}\label{eq_FluxQuant}
\varphi_a-\varphi_b+\varphi_c=-2 \pi f\ ,
\end{equation}
where we have defined $f=\Phi/\Phi_0$.
The total Josephson energy of the ring is given by the phase-dependent
expression $U=\sum_i E_\mathrm{J}^{i}(1-\cos\varphi_i)$. Combining this
relation with the flux quantisation condition in \eref{eq_FluxQuant} the
Josephson energy can be written as
\begin{equation}\label{eq_JEnergy}
U=
E_\mathrm{J} [2+\alpha-\cos\varphi_\mathrm{a}-\cos\varphi_\mathrm{b}-\alpha\cos(2 \pi
f+\varphi_\mathrm{a}-\varphi_\mathrm{b})]\ .
\end{equation}
The potential $U$ is plotted in figure~\ref{fig_DelftQubit} (b) for $\alpha=0.75$ and
$f=0.495$, a typical operation point of the Delft qubit. The plot shows a periodic structure of two nearby minima. These two minima, indicated by L and R,  fulfill the condition
$\varphi_\mathrm{a}=-\varphi_\mathrm{b}\equiv\varphi$ and correspond to situations
in which the Josephson current in the loop has opposite signs. Due to the
periodicity of the potential, all other minima are equivalent to L and R. If
the magnetic flux is tuned to $f=\frac{1}{2}$, the flux point usually called ``sweet spot'', the two minima are equal,
$U_\mathrm{min}=2E_\mathrm{J}\left(1-\frac{1}{\alpha}\right)$, and they are
situated at $\varphi_\mathrm{L/R}=\mp \arccos(1/2\alpha)$.
Small deviations $\delta f=f-\frac{1}{2}$ from this point yield a shift of the minima by
$\delta\varphi=-2\pi\,\delta f\,(2\alpha^2-1)/(4\alpha^2-1)$, such that
$\varphi_\mathrm{L/R} = \mp \arccos(1/2\alpha) +\delta\varphi$. Consequently, the
potential becomes asymmetric as indicated in  figure~\ref{fig_DelftQubit}  (c). For values  $\alpha \leq 1/2$ the two minima would merge into a single minimum; in the following we will hence always assume $\alpha > 1/2$.

\begin{figure}[t]
\begin{center}
\includegraphics[width=0.6\linewidth]{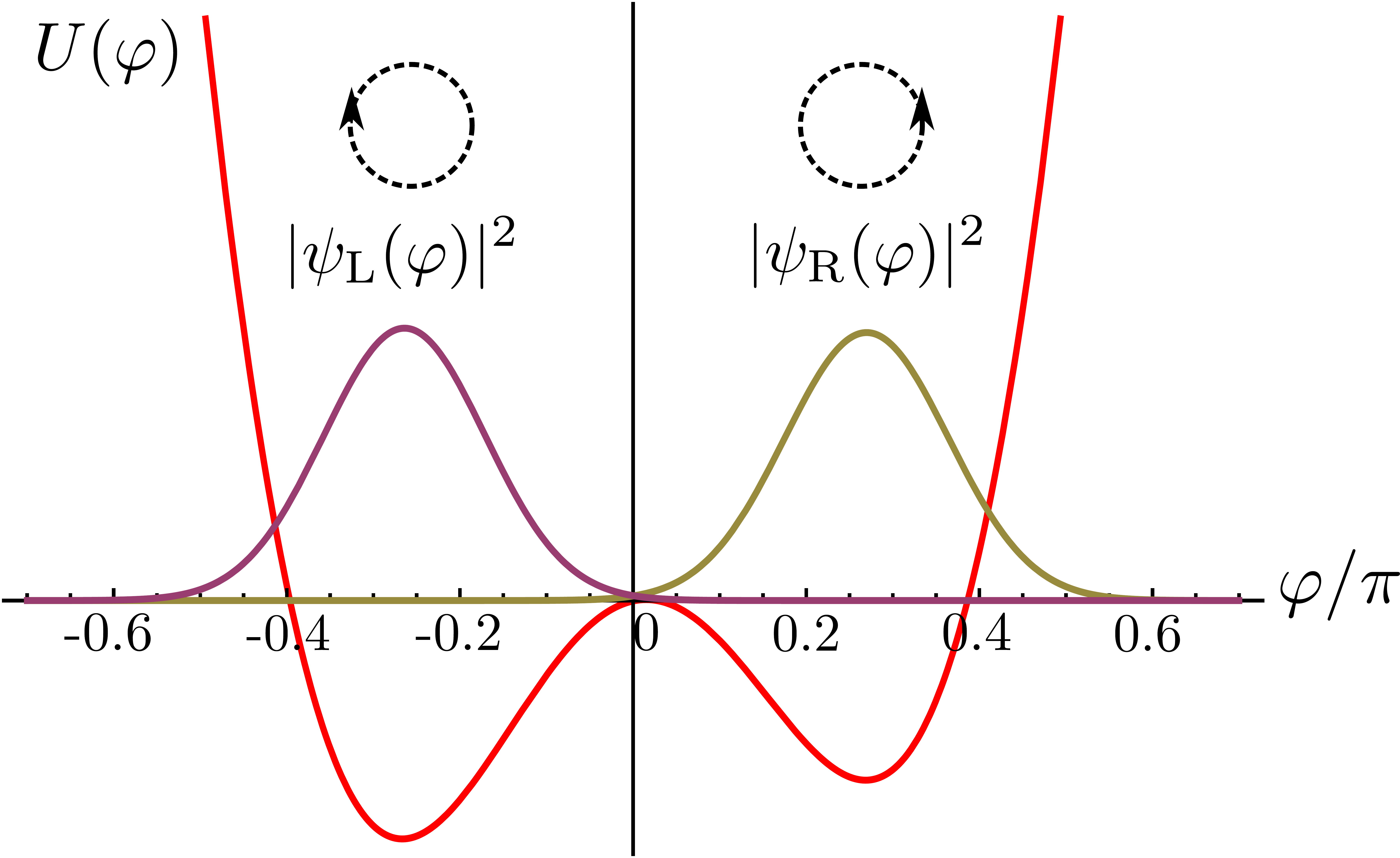}
\caption{Cut through the potential of the SQUID along $\varphi=\varphi_\mathrm{a}=-\varphi_\mathrm{b}$, with the two approximated qubit states
$|\psi_\mathrm{L}\rangle$ and $|\psi_\mathrm{R}\rangle$ in the phase 
representation for $\alpha = 0.75$ and $f = 0.495$. }\label{fig_approximation}
\end{center}
\end{figure}

 We are interested in the quantum properties of this system and therefore use
 a Hamiltonian description taking
 $\varphi$ as general coordinate. The dynamics of the system is provided by
 the fact that each of the junctions adds a small electrical
 capacitance $C$. In fact the conjugate momentum to $\varphi$ is given by 
 the number of Cooper pairs $N=-i\hbar\;\partial/\partial\varphi$, which charge
 the capacitances. We arrive
 at the Hamiltonian
\begin{equation}\label{Hamiltonian}
H_\mathrm{qubit}=-4 E_\mathrm{C}\frac{\partial^2}{\partial\varphi^2}+E_\mathrm{J}\bigg[2+\alpha-2\cos(\varphi)-\alpha\cos(2\pi f +2\varphi)
\bigg]\ ;
\end{equation}
the first term takes account for the charging energy $E_\mathrm{C} =
e^2/2C_\Sigma$, where $C_\Sigma$ combines the capacitive effects of the three junctions,  and the second term is the potential energy $U$ as given in
\eref{eq_JEnergy} with $\varphi=\varphi_\mathrm{a}=-\varphi_\mathrm{b}$.
The low-energy physics of this system can be described by the two metastable 
states $|\psi_\mathrm{L}\rangle$ and $|\psi_\mathrm{R}\rangle$, corresponding to the
ground states of the local minima of the potential as shown in
figure~\ref{fig_DelftQubit}  (c). They will serve as the two qubit states in
the following. In the vicinity of the local minima, the Hamiltonian can be
approximated using $U(\varphi)\approx U(\varphi_\mathrm{L/R})+ E_\mathrm{J}\left[\cos(\varphi_\mathrm{L/R})+2\alpha\cos(2\pi f +2\varphi_\mathrm{L/R})\right](\varphi-\varphi_\mathrm{L/R})^2$ for $\varphi\approx\varphi_\mathrm{L/R}$. The qubit states are then given by the oscillator ground states
\begin{equation}\label{eq_wavefunctions}
\langle\varphi|\psi_\mathrm{L/R}\rangle=\left(\frac{\lambda_\mathrm{L/R}}{\pi}\right)^{1/4}\mathrm{exp}\left\{-\frac{\lambda_\mathrm{L/R}(\varphi-\varphi_\mathrm{L/R})^2}{2}\right\}
\end{equation}
with the inverse of the variance
\begin{eqnarray}\label{lambdaapprox}
 \lambda_\mathrm{L/R}&= \frac{E_\mathrm{J}}{2E_\mathrm{C}}[\cos{(\varphi_\mathrm{L/R})}+2\alpha\cos(2\pi
  f +2\varphi_\mathrm{L/R})]\nonumber\\
  &\approx\frac{E_\mathrm{J}}{2E_\mathrm{C}}
  \left[\frac{4 \alpha^2 -1}{2\alpha} \mp \frac{\pi (1+ 2
  \alpha^2) \delta f}{\alpha\sqrt{4 \alpha^2-1}}  \right]
\end{eqnarray}
These states are shown in figure~\ref{fig_approximation}
together with the qubit potential. They are coupled through possible quantum tunnelling through the potential barrier between the two minima, of which the height that  depends on the values of the asymmetry parameter, $\alpha$, is tuneable via the Josephson energy of the junction c.~\footnote{This
can be done for instance by replacing junction c by an additional 2-junction
SQUID with a separately tuneable flux~\cite{Mooij99,Makhlin01}.} However, as soon as the flux deviates
from the value $\Phi=\Phi_0/2$ the qubit eigenstates occur to be
well-localised in the potential wells, coupling between the two states is negligibly small and they are hence approximately given by
$|\psi_\mathrm{L}\rangle$ and $|\psi_\mathrm{R}\rangle$.

\subsection{Microscopic model of the Josephson junctions}

In order to calculate heat currents flowing through the SQUID it is important
to consider the \textit{microscopic} model of the three junctions.  The
microscopic Hamiltonian for two superconducting arms, which we here choose to
be $l=1,2$, connected by a tunnel contact is given by
\begin{eqnarray}\label{eq_Microscopic} H_\mathrm{junction} & = &
\sum_{l=1,2}\sum_{k,\sigma}\xi_{l,k\sigma}c^\dagger_{l,k\sigma}c^{}_{l,k\sigma}
-\sum_{l=1,2}\sum_{k}\left(\Delta_lc^\dagger_{l,k\uparrow}c^\dagger_{l,-k\downarrow}+\mathrm{H.c.}\right)\nonumber\\
&&+\sum_{k,k',\sigma} \left(
V^{12}_{kk'}c^\dagger_{1,k\sigma}c^{}_{2,k'\sigma}+\mathrm{H.c.} \right)
\end{eqnarray} 
where $\xi_{l,k\sigma}=\varepsilon_{l,k\sigma}-\mu_0$ is the electron energy relative 
to the chemical potential which we take equal for all electrodes, $\mu_l=\mu_0, \forall l$.
The Hamiltonian for the other two junctions of the SQUID is
found equivalently.  The creation (annihilation) operators for electrons in
reservoir $l$, with momentum $k$ and spin $\sigma=\uparrow,\downarrow$ are
given by $c^\dagger_{l,k\sigma}(c^{}_{l,k\sigma})$.  The two superconductors
are kept at different temperatures $T_l$ and have a superconducting gap $\Delta_l$, which we here assume to be
independent of $k$.  The gap is characterized by its absolute value
$|\Delta_l|$ and the phase $\phi_{l}$.  This phase enters the heat current
across the junction only in phase differences $\varphi_{12}$ across a
junction, with $\varphi_{12}=\phi_1-\phi_2$.  For the SQUID model considered in this manuscript we
have $\varphi_{12}=\varphi_\mathrm{a}$, etc.  The temperature dependence of
the magnitude of the  superconducting gap is approximately given by
$|\Delta_l(T_l)|\approx \Delta_0\sqrt{1-T_l/T_\mathrm{crit}}$ with $\Delta_0\simeq k_\mathrm{B}
T_\mathrm{crit}$ the gap of the superconductor at zero temperature and
$T_\mathrm{crit}$ the critical
temperature.  Here and in the
following, we assume that all the superconductors are built from the same
material with equivalent geometries, such that they share $T_\mathrm{crit}$ and $\Delta_0$.  Tunnelling between the two superconductors 1 and 2 occurs with the
tunnelling amplitude $V^{12}_{kk'}$.  The resistance of the junction
connecting reservoirs 1 and 2 is related to the normal conducting density of
states of the reservoirs at the Fermi level (including
spin), $N^0_l$,  and the tunneling amplitude; the inverse resistance is given by
$R_{12}^{-1}= \pi e^2 N_1^0 N_2^0 |V^{12}|^2/\hbar$.

\subsection{Heat currents in superconducting links}

We are in the following interested in the heat currents flowing through the
junctions in the SQUID, when the arms between the junctions are kept at
different temperatures, as sketched in figure~\ref{fig_DelftQubit} (a). The ring is supposed to be large enough such that the arms between the junctions are larger than the quasiparticle coherence length and we can therefore model the arms as quasiparticle reservoirs and treat the heat current through the junctions separately. Note, that the phase differences across the junctions are related through the superconducting fluxoid quantization given in (\ref{eq_FluxQuant}).
The heat current in electrode $l$ is defined as the flow of energy with
respect to the electrochemical potential of electrode $l=1,2,3$, 
\begin{equation}
\dot{Q}^l= \frac{d}{dt}\langle H_l\rangle = -\frac{i}{\hbar}\langle\left[H, H_l\right]\rangle\ ,
\end{equation}
where $H_l$ is given by the first line of~(\ref{eq_Microscopic}).
We are subsequently interested in the weak
tunnel coupling regime~\cite{Maki65,Zhao04}, see the appendix for a
detailed derivation of the heat current. The heat current through a junction connecting
reservoir $l$ and $m$ due to a difference in temperature, $T_l\neq T_m$, with
$l,m=1,2,3$, can then be divided into a  pure quasiparticle contribution to
the heat current, $\dot\Qqp^l(T_l,T_m)$, and an interference contribution due to an
interplay between quasiparticles and the Cooper pair condensate, $\dot\Qint^l(T_l,T_m)$, namely
\begin{equation}\label{eq_dQdt}
\dot{Q}^l(T_l,T_m)=\dot{Q}^l_{\mathrm{qp}}(T_l,T_m)-\dot{Q}^l_{\mathrm{int}}(T_l,T_m)\cos{\varphi_{lm}}\ .
\end{equation}
We find the pure quasiparticle contribution to the heat current to be
\begin{equation}\label{eq_Qqp}
  \dot{Q}^l_{\mathrm{qp}}(T_l,T_m)=\frac{2}{e^2R_{lm}}\int_{|\Delta_\mathrm{max}|}^{\infty}d\omega\,\omega^3\frac{f_l(\omega)-f_m(\omega)}{\sqrt{\omega^2-|\Delta_l|^2}
\sqrt{\omega^2-|\Delta_m|^2}}\ ,
\end{equation}
where $f_l(\omega)=\left[1
+\exp(\omega/k_\mathrm{B}T_l)\right]^{-1}$ is the Fermi function of electrode
$l$ and $|\Delta_\mathrm{max}|=\max\left\{|\Delta_l|,|\Delta_m|\right\}$. The interference contribution to the heat current due to  the interplay between quasiparticles and the Cooper pair condensate  depends on the phase
difference $\varphi_{lm}$ of the superconducting condensates and yields
\begin{equation}\label{eq_Qint}
\dot{Q}^l_{\mathrm{int}}(T_l,T_m)=
\frac{2}{e^2R_{lm}}\int_{|\Delta_\mathrm{max}|}^{\infty}
\!\!\!d\omega\,\omega
|\Delta_l\Delta_m|\frac{f_l(\omega)-f_m(\omega)}{
\sqrt{\omega^2-|\Delta_l|^2}\sqrt{\omega^2-|\Delta_m|^2}}\ .
\end{equation} 
We have $T_l, T_m\lesssim |\Delta_0|/k_\mathrm{B}$ and the square root
terms are changing faster than the other factors in the integrals of
\eref{eq_Qqp} and \eref{eq_Qint}. The magnitude of the heat currents can
then be estimated as
\begin{eqnarray}\label{eq:estimate}
  \dot\Qqp^l(T_l,T_m)&\simeq \dot\Qint^l(T_l,T_m)\simeq \dot Q_\mathrm{typ}\\
  &= \frac{|\Delta_\mathrm{max}|^2}{e^2 R_{lm}}
  K(|\Delta_\mathrm{min}|/|\Delta_\mathrm{max}|)
  [e^{-|\Delta_\mathrm{max}|/k_\mathrm{B} T_l} -
  e^{-|\Delta_\mathrm{max}|/k_\mathrm{B} T_m} ] \nonumber\ ,
\end{eqnarray}
with $K(k)= \int_0^{\pi/2} (1- k^2 \sin^2 \phi)^{-1/2} d\phi$ the complete
elliptic integral of the first kind, and $|\Delta_\mathrm{min}|$ the superconducting gap at the larger temperature.  The elliptic integral is a monotonously
increasing function which starts at $\pi/2$ for small arguments and has a
logarithmic divergence with $K(1-\epsilon^2) \sim \ln 1/\epsilon$ when $k$
approaches 1.  Since the contribution of the integrands of (\ref{eq_Qqp}) and
(\ref{eq_Qint}) have a maximum for $\omega$ being in the vicinity of the
superconducting gap $|\Delta_\mathrm{max}|$, the quasiparticle and the
interference contributions to the heat current are generally of the same order
of magnitude.

Before discussing the sensitivities of the heat currents to the qubit state,
we here want to briefly give an estimate of the order of magnitude of the
heat currents for the limits of small and large temperature differences.  In
the case of a \textit{small temperature difference}, $\delta T
\equiv|T_l-T_m|\ll T_l, T_m$, we obtain from \eref{eq:estimate} the typical
value $ \dot Q_\mathrm{typ}\simeq |\Delta_\mathrm{max}|^3 K[1- \delta T/2
(T_\mathrm{crit} - T)] e^{-|\Delta_\mathrm{max}|/k_\mathrm{B} T}\delta T /(e^2
R_{lm} k_\mathrm{B} T^2) $ of the heat current.  Assuming furthermore that
$\delta T \ll T_\mathrm{crit} -T $, we obtain the estimate $ \dot
Q_\mathrm{typ}\ \simeq |\Delta_\mathrm{max}|^3 \ln [ (T_\mathrm{crit} -T)
/T_\mathrm{cut}] e^{-|\Delta_\mathrm{max}|/k_\mathrm{B} T}\delta T /(e^2
R_{lm} k_\mathrm{B} T^2) $.  The tunneling approximation gives a cutoff
temperature $T_\mathrm{cut} = \delta T$ which leads to a logarithmic
divergence of the heat current for small temperature gradients as already
pointed out in \cite{Maki65}.  However, as shown in \cite{Zhao04} this is an
artifact of the tunneling approximation which fails to take properly into account a resonance in the
density of states due to a weakly bound Andreev state.
The resonance introduces a new cutoff at the scale $T_\mathrm{cut} = D
\Delta_0 \sin^2 (\varphi_{lm}/2)/k_B$ with $D$ the transparency of the
tunneling barrier.

 In contrast, in the case of a
\textit{large temperature difference}, we have that $T_\mathrm{min} \ll
T_\mathrm{max}$.  Since in this case $T_\mathrm{min}$ is hence also always much smaller than $T_\mathrm{crit}$, the heat current only depends on
$T_\mathrm{max}$ and we obtain the estimate $ \dot Q_\mathrm{typ}\simeq
\Delta_0^2 e^{-\Delta_0/k_\mathrm{B} T_\mathrm{max}} K(\sqrt{1-
T_\mathrm{max}/T_\mathrm{crit}})/e^2 R_{lm} $.  If we additionally have that
$T_\mathrm{max} \lesssim T_\mathrm{crit}$, the elliptic integral is of order one and the
estimate simply reads $ \dot Q_\mathrm{typ}\simeq \Delta_0^2
e^{-\Delta_0/k_\mathrm{B} T_\mathrm{max}}/e^2 R_{lm} $.

Thermal currents in a system similar to the one we study here were measured in the experiment by Giazotto and Martinez-Perez reported in
\cite{Giazotto12a}. If we use these same experimental values for an estimate, we have $\Delta_0\simeq200\,\mu$eV and
$R\simeq1\,$k$\Omega$.  For $T=0.1 T_\mathrm{crit}$ and large temperature
gradient we obtain $\dot{Q}_\mathrm{typ}\simeq10^{-11}$ W, while for a small temperature
gradient we obtain the estimate $\dot{Q}_\mathrm{typ}\simeq (\delta T/T) 10^{-14} \,$W
(assuming that the logarithm is of order one).

\section{Qubit-state sensitive heat currents}

In the following, we want to investigate the sensitivity of the heat current to the state of the persistent-current qubit realised by the three-junction SQUID introduced before. We therefore propose to study  the difference between the heat currents compared to the sum of the two currents for the qubit being in the state $|\psi_\mathrm{L}\rangle$ or $|\psi_\mathrm{R}\rangle$, characterising the sensitivity,
\begin{equation}
s_l=\frac{\dot{Q}_\mathrm{R}^{l}-\dot{Q}^{l}_\mathrm{L}}{\dot{Q}^{l}_\mathrm{R}+\dot{Q}^{l}_\mathrm{L}}\;\;\;\;\; \mathrm{with} \;\;\;\;\; \dot{Q}^{l}_\mathrm{L/R}=\langle\psi_\mathrm{L/R}|\dot{Q}^l|\psi_\mathrm{L/R}\rangle
\end{equation}
The expectation values are obtained from the usual integral over $\varphi$ of the product of the heat currents given in (\ref{eq_dQdt}) with the wave functions of (\ref{eq_wavefunctions}). We evaluate the heat currents in each electrode due to a temperature gradient induced by $T_1=T_2<T_3$. This yields heat currents in electrodes 1 and 2 given by the heat flow through the junction with electrode 3 only, while the heat current in electrode 3 has two contributions. To simplify the notation, we now take as a reference the heat current into electrode 1, with $\dot{Q}_\mathrm{int}\equiv \dot{Q}^{1}_\mathrm{int}(T_1,T_3)$ and $\dot{Q}_\mathrm{qp}\equiv \dot{Q}^{1}_\mathrm{qp}(T_1,T_3)$. The sensitivities then take the simple form 
\begin{eqnarray}\label{eq_sensitivities}
s_1&=& \frac{\dot{Q}_\mathrm{int}\big(C_\mathrm{L}-C_\mathrm{R}\big)}{2\dot{Q}_\mathrm{qp}-\dot{Q}_\mathrm{int}\big(C_\mathrm{L}+C_\mathrm{R}\big)}\nonumber\\
s_2&=&\frac{\dot{Q}_\mathrm{int}\big(D_\mathrm{L}-D_\mathrm{R}\big)}{2\dot{Q}_\mathrm{qp}-\dot{Q}_\mathrm{int}\big(D_\mathrm{L}+D_\mathrm{R}\big)}\nonumber\\
s_3&=&\frac{\dot{Q}_\mathrm{int}\big(\alpha(D_\mathrm{L}-D_\mathrm{R})+(C_\mathrm{L}-C_\mathrm{R})\big)}{2\dot{Q}_\mathrm{qp}(1+\alpha)-\dot{Q}_\mathrm{int}\big(\alpha(D_\mathrm{L}+D_\mathrm{R})+(C_\mathrm{L}+C_\mathrm{R})\big)}\ .
\end{eqnarray}
For a short notation and assuming the two qubit states to be well localized, we here define the phase-dependent factors
\begin{eqnarray}
C_\mathrm{L/R}&=&\langle \psi_\mathrm{L/R} | \cos (\varphi_\mathrm{a}) | \psi_\mathrm{L/R}\rangle =
\cos(\varphi_\mathrm{L/R})e^{-1/\left(4\lambda_\mathrm{L/R}\right)},\nonumber\\
D_\mathrm{L/R}&=&\langle \psi_\mathrm{L/R} | \cos (\varphi_\mathrm{c}) | \psi_\mathrm{L/R}\rangle =
\cos(2\varphi_\mathrm{L/R}+2\pi f)e^{-1/\lambda_\mathrm{L/R}}.
\end{eqnarray}
We also used the generalized Ambegaokar-Baratoff relations~\cite{Giazotto05,Tirelli08} in order to relate the heat currents $\dot{Q}^{(i)}$ through the junctions $i=\mathrm{b,c}$ to each other, when $T_1=T_2$. The heat currents through the \textit{different junctions} are furthermore related to the heat currents $\dot{Q}^{l}$ \textit{into the different reservoirs}, $l=1,2,3$, by $\dot{Q}^{i=\mathrm{b}}\equiv\dot{Q}^{1}$, $\dot{Q}^{i=\mathrm{c}}\equiv\dot{Q}^{2}$  and hence $\dot{Q}^{3}=-\dot{Q}^{i=\mathrm{b}}-\dot{Q}^{i=\mathrm{c}}$.  By comparing the separate quasi-particle and interference components of these heat currents, see (\ref{eq_Qqp}) and (\ref{eq_Qint}), we then find 
\begin{equation}
\frac{\dot{Q}^{\mathrm{c}}_\mathrm{int}}{\dot{Q}^{\mathrm{b}}_\mathrm{int}}=\frac{\dot{Q}^{\mathrm{c}}_\mathrm{qp}}{\dot{Q}^{\mathrm{b}}_\mathrm{qp}}=\frac{I^{\mathrm{b}}_\mathrm{crit}}{I^{\mathrm{c}}_\mathrm{crit}}=\frac{R_\mathrm{c}}{R_\mathrm{b}}=\frac{R_{23}}{R_{13}}=\alpha\ .
\end{equation}
This finally leads to the compact expressions in (\ref{eq_sensitivities}).
\begin{figure}[t]
\begin{center}
\includegraphics[width=0.60\linewidth]{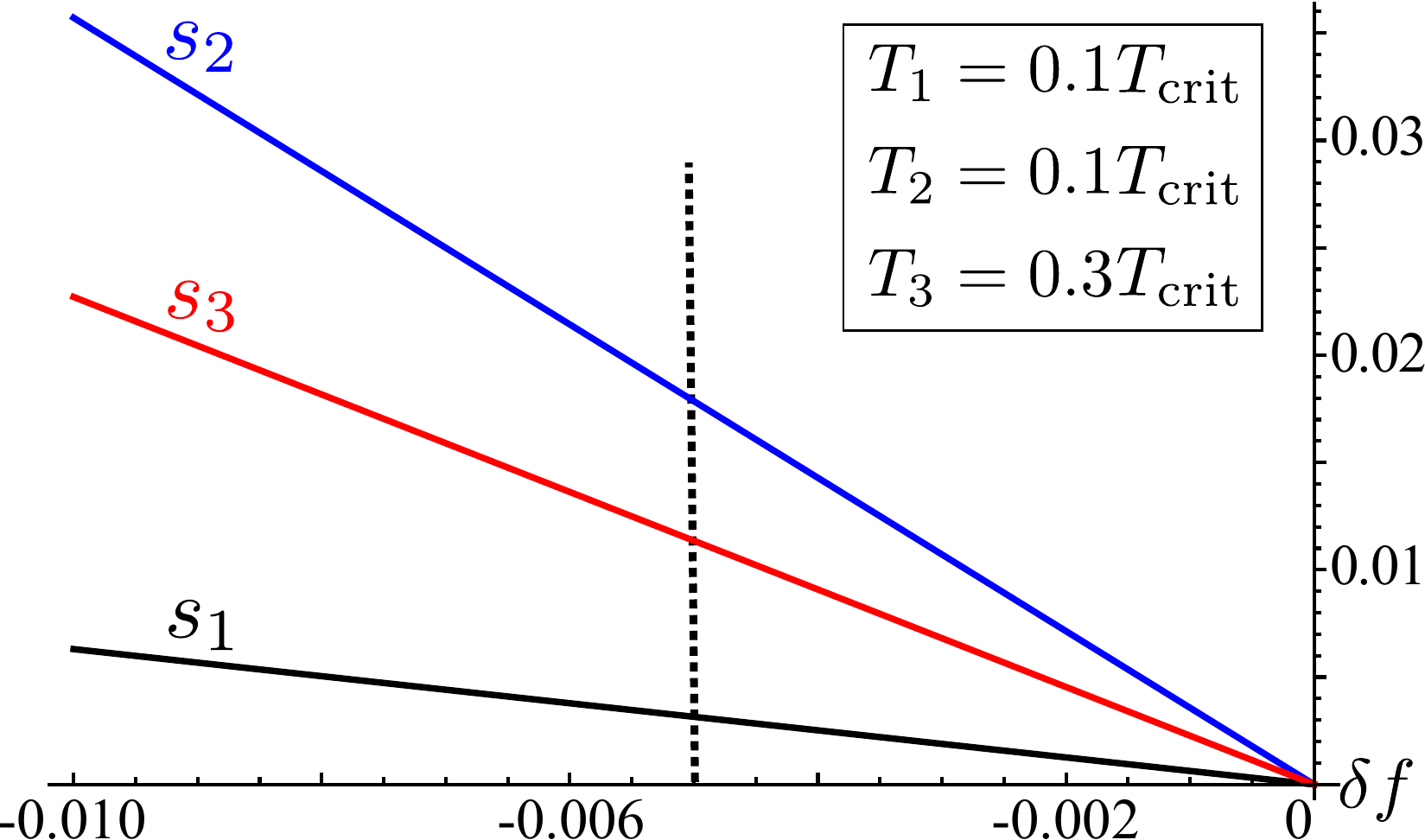}
\caption{Plot of the sensitivities, $s_l$, for the three electrodes $l=1,2,3$ as a function of the flux enclosed in the loop, for $\alpha=0.75$ and $E_\mathrm{J}/E_\mathrm{C}\approx80$. The vertical dotted line indicates the flux value of the Delft qubit ``operation point''~\cite{Mooij99}.}
\label{fig_SemiQuant}
\end{center}
\end{figure}
The results for these three sensitivities as a function of the flux, in the vicinity of the sweet spot and the operation point of the Delft qubit, are shown in figure~\ref{fig_SemiQuant}.
The sensitivity of the heat currents to the qubit state hence yields a
possible measure of the latter. The heat currents in electrodes 2 and 3 are
most sensitive to the qubit state with a sensitivity of about $2\%$ at the ``operation point'', $f=0.495$~\cite{Mooij99}. The plot in
figure~\ref{fig_SemiQuant} shows a dependence of the sensitivities as a function of the magnetic flux penetrating the SQUID which is very close to a linear function. The slopes of the latter depend on the specific realisation of the qubit, namely on the ratio $\alpha$, on the electrode temperatures $T_l$ and the applied thermal gradient, as well as on the ratio of Josephson and charging energy. This is shown in the approximate result for the heat currents for small deviations $\delta f$ from the  ``sweet spot'' $f=\frac{1}{2}$, 
$s_l\approx m_l(\alpha)\delta f$ with the respective slope $m_l$. 
\begin{figure}[t]
\begin{center}
\includegraphics[width=0.60\linewidth]{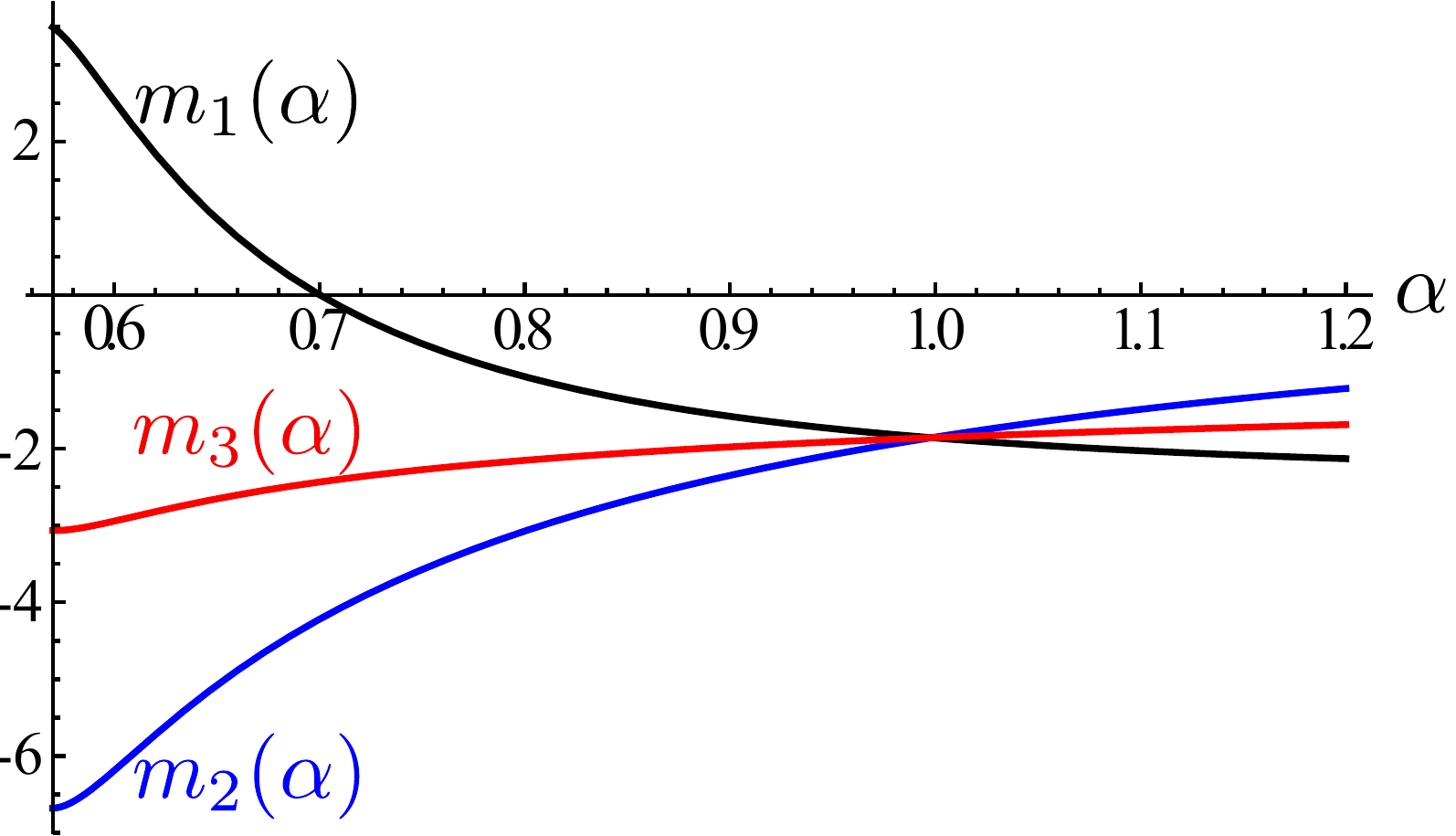}
\caption{Coefficient of the linear expansion of the heat currents of the electrodes as a function of $\alpha$ for $T_1=T_2=0.1 T_\mathrm{crit}$, $T_3=0.3 T_\mathrm{crit}$, and $E_\mathrm{J}/E_\mathrm{C}\approx80$.}
\label{fig_m_alpha}
\end{center}
\end{figure}
The rather complex explicit analytic form of the slopes of the three sensitivities are given
in~\ref{app_slopes} and they are shown in figure~\ref{fig_m_alpha} as a function of the ratio $\alpha$, which is tuneable in the experiment.  We find that the slopes of the sensitivities do in general not need to have the same sign. The slope with the largest absolute value is the one obtained from the heat currents into reservoir 2. This is related to the fact that the heat currents into this reservoir flow uniquely through the junction with the weakest Josephson coupling, namely junction c, which has consequently the largest phase difference and is most sensitive to the qubit state. While for the working point of the Delft qubit, that is at $\alpha\approx0.75$, the slope of $s_2$ has already a rather large value, this value can be improved by lowering $\alpha$. Note however that with $\alpha$ approaching 0.5 the two valleys of the potential get closer and the qubit states are not well defined any more. Equivalently for $\alpha>1$ the SQUID can not be used as a qubit any longer.

\section{Impact of temperature gradients on the qubit dephasing}

\begin{figure}[t]
\begin{center}
\includegraphics[width=0.60\linewidth]{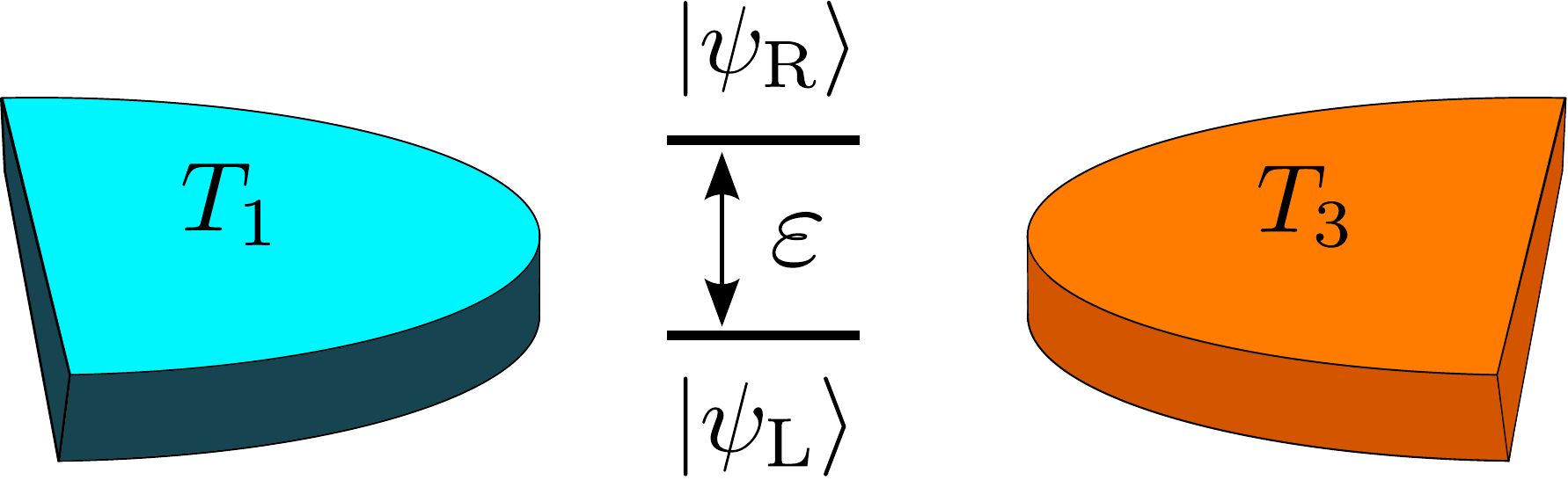}
\caption{Model of the two-level system with level-spacing $\epsilon$, tunnel-coupled to two quasi-particle baths at different temperatures $T_1$ and $T_3$.}
\label{fig_qubitmodel}
\end{center}
\end{figure}

After having demonstrated the sensitivity of the heat currents to the state of the qubit, the aim of this section is to study the impact of a temperature gradient - and the resulting heat current - on the coherence properties of the qubit. Our interest in this point is twofold: on one hand we want to find out the behavior of the qubit state under measurement, on the other hand we are interested in the impact of accidental temperature gradients on the dephasing of the qubit. We therefore consider the two-level system, defined by the states $|\psi_\mathrm{L}\rangle$ and  $|\psi_\mathrm{R}\rangle$, namely the qubit states obtained from the low-energy physics of the SQUID, in contact with two heat baths, resulting in the model Hamiltonian
\begin{eqnarray}\label{eq:toy}
  H_\mathrm{toy} & = & -\frac{\varepsilon}{2}\tau^3- \frac{w}2\tau^1
  +\sum_{l=1,3}\sum_{k,\sigma}
  (\varepsilon_{l,k} -\mu_l) a_{l,k\sigma}^{\dagger}
  a_{l,k\sigma}^{}
  \nonumber\\
  && +\sum_{k,q,\sigma} \Bigl[
  a_{1,k \sigma}^\dagger a_{3,q\sigma}^{} (V_0\tau^0
  + V_3
  \tau^3) + \mathrm{H.c.} \Bigr].
\end{eqnarray}
The model is depicted in figure~\ref{fig_qubitmodel}. Here, the matrices
$\tau^j, j=0,1,3$ are Pauli matrices in the qubit space. The level 
splitting between the qubit states is given by $\varepsilon$ and weak 
coupling between them is
denoted by $w$.   The creation (annihilation) operators of particles with
momentum $k$ and spin $\sigma$ in lead $l$ are given by
$a^{\dagger}_{l,k\sigma}(a^{}_{l,k\sigma})$. 

In the simplified model \eref{eq:toy}, we do not explicitly take into account the
three superconducting leads with the heat currents, which depend on all
three phase differences, but rather discuss a simplified microscopic model,
which involves only two leads.  The idea is to set the density of states and
tunnelling matrix elements such as to reproduce the correct macroscopic thermal
current between the reservoirs at temperature $T_1=T_2$ and $T_3$ in the three
lead setup. We expect that such a procedure, while being inaccurate for certain
microscopic details, will correctly incorporate the effects of the 
phase-dependent thermal currents on the qubit.
In linear response, the Hamiltonian (\ref{eq:toy}) leads to a heat current,
\begin{equation}\label{eq:thermal}
  \dot Q^\mathrm{toy}_\mathrm{L/R} 
  = \frac{\pi}{\hbar}\int_{-\infty}^{\infty} d\omega\,\omega
   (|V_0|^2 \pm |V_3|^2)
   N_{1}(\omega) N_{3}(\omega) [ f_1(\omega)
  -f_3(\omega) ]
\end{equation}
with $N_l(\omega)$ the density of states of the electrons in lead $l$
(including spin).
If we set the parameters
\begin{eqnarray}\label{eq:v_and_rho}
  |V_0|^2 &=& \frac12|V^{13}|^2 \left[ 2\alpha+2 - C_\mathrm{R}- C_\mathrm{L} -\alpha\left( D_\mathrm{R} + D_\mathrm{L}\right)\right]\nonumber\\
  |V_3|^2 & = & \frac12 |V^{13}|^2\left[C_\mathrm{R}  - C_\mathrm{L}+ \alpha\left(D_\mathrm{R} - D_\mathrm{L}\right)\right] \nonumber\\
  N_{l}(\omega)&=& N_l^0  \frac{|\Delta_l|}{\sqrt{\omega^2-|\Delta_l|^2}}
  \theta(\omega^2-|\Delta_l|^2),
\end{eqnarray}
with $\theta(x)$ the unit-step function, we achieve the goal of reproducing the
correct qubit-state dependent heat current with
$\dot Q^\mathrm{toy} = \dot Q^3 =
-\dot Q^1 - \dot Q^2$;  
here and below, we assume the magnitude of the quasiparticle and interference parts of the heat current to be equal. 

We are now interested in the dynamics of the qubit state depending on the qubit-state sensitive heat current induced by the temperature gradient. Starting from the full system's density matrix, we therefore trace out the lead degrees of freedom and write down a master equation for the reduced density matrix of the qubit, $\rho(t)$.
If we write the density matrix of the qubit as $\rho(t)=\frac{1}{2}\left[\mathds{1}+\bm{\tau}\cdot \bm{S}(t)\right]$ with $\bm S(t) = \tr [ \rho(t) \bm \tau ]=(\rho_\mathrm{LR}(t)+\rho_\mathrm{RL}(t), i(\rho_\mathrm{LR}(t)-\rho_\mathrm{RL}(t)), \rho_\mathrm{LL}(t)-\rho_\mathrm{RR}(t))^\mathrm{T}$,
we obtain the Pauli rate equation
\begin{equation}
\dot{\bm S}(t)=
\bm{S}(t) \times \bm{h}-\gamma(S_1(t),S_2(t),0)^\mathrm{T}\ .
\end{equation}
This equation contains a precession around a pseudo-magnetic field, $\bm h= (w,0,\varepsilon)^\mathrm{T}$, determined by the qubit properties, and
a relaxation of the coherences of the reduced density matrix with the rate $\gamma$, while the diagonal elements, namely the occupations of the qubit states,  do not decay. This is also appreciable from the solution of the master equation, which for large detuning $\varepsilon \gg w$ with respect to the weak tunneling between the qubit states, is given by 
\begin{eqnarray}\label{eq_master_solution}
\rho_\mathrm{LL}(t)\approx\rho_\mathrm{LL}(0), & \  \ \ \ \ \ &
\rho_\mathrm{RR}(t)\approx\rho_\mathrm{RR}(0),\nonumber\\
\rho_\mathrm{LR}(t)\approx\rho_\mathrm{LR}(0)e^{-(\gamma+i\varepsilon) t}, & \  \ \ \ \ \ & 
\rho_\mathrm{RL}(t)\approx\rho_\mathrm{RL}(0)e^{-(\gamma-i\varepsilon) t}\ .
\end{eqnarray}
The value of the dephasing rate $\gamma$ reads,
\begin{eqnarray}
  \gamma&= \frac{4\pi |V_3|^2 N_1^0N_3^0}\hbar
  \int_{|\Delta_\mathrm{max}|}^\infty d\omega\
\omega^2
\frac{[1-f_1(\omega)]f_3(\omega)+[1-f_3(\omega)]f_1(\omega)}{\sqrt{\omega^2-|\Delta_1|^2}\sqrt{\omega^2-|\Delta_3|^2}}.
\end{eqnarray}
Importantly, this rate equals zero, if the sensitivity of the heat current to
the qubit state vanishes and hence $|V_3|^2\propto C_\mathrm{R}  - C_\mathrm{L}+ \alpha\left(D_\mathrm{R} -
D_\mathrm{L}\right) =0$. Note that this means that the temperature gradient leads to dephasing only when the qubit is tuned away from the sweet spot. Indeed, it is possible to conclude that the qubit-state sensitivity of the heat current represents a measurement process which reflects in the time-dependent solution of the master equation given in (\ref{eq_master_solution}).

The dephasing rate is connected to fluctuations in the electronic subsystem
which drive the qubit.  In equilibrium, we would expect a
fluctuation-dissipation relation to hold which relates the fluctuations to the
response coefficient of the system. Naturally, this is not true in the non-equilibrium situation studied here. It is however interesting to compare the response of the system to the temperature gradient, namely the heat current depending on the qubit states, to the related dephasing rate. We therefore introduce the dimensionless ratio $r=
|\Delta_\mathrm{max}|\mathrm{\gamma}/ |\dot Q^3_\mathrm{L} - \dot Q^3_\mathrm{R}
|$.  As above, we specialise to the case when $T_1,
T_3 \lesssim |\Delta_\mathrm{max}|/k_\mathrm{B}$. With a similar
calculation as the one following~\eref{eq_Qint}, we obtain the
estimate
\begin{equation}\label{eq:r}
  r\simeq \coth \left(\frac{|\Delta_\mathrm{max}| |T_1 - T_3|}{2k_\mathrm{B} T_1 T_3}
  \right)\ .
\end{equation}
This means that $r$ is universal with respect to microscopic details like the normal-state
resistance $R_{13}$ or the phase difference $\varphi_j$ across the junctions, and only depends on thermodynamical quantities like the temperatures $T_1,
T_3$ and the gap $\Delta_0$.
We see that for small temperature differences, $\delta T = T_1 -T_3 \ll
T_1,T_3$, this ratio becomes $r\simeq  k_\mathrm{B}T^2/\left( |\Delta_\mathrm{max}| \delta T\right)$. 

The
dephasing time $\tau_\phi = \gamma^{-1}$ is in this case given by
\begin{eqnarray}\label{dephtimedT}
  \tau_\phi &\simeq \frac{\Delta_0^2(1-T/T_\mathrm{crit}) \delta T}
  { k_\mathrm{B}T^2 | \dot Q^3_\mathrm{L} - \dot Q^3_\mathrm{R}
  |}\\
  &\simeq\frac{e^2R_{13} e^{\Delta_\mathrm{max}/k_\mathrm{B}T}}
  {\Delta_\mathrm{max}\ln(T_\mathrm{crit}/ T_\mathrm{cut} )
  [C_\mathrm{L}-C_\mathrm{R}+\alpha(D_\mathrm{L}-D_\mathrm{R})]}.
  \end{eqnarray}
\begin{figure}[t]
\begin{center}
\includegraphics[width=13.cm]{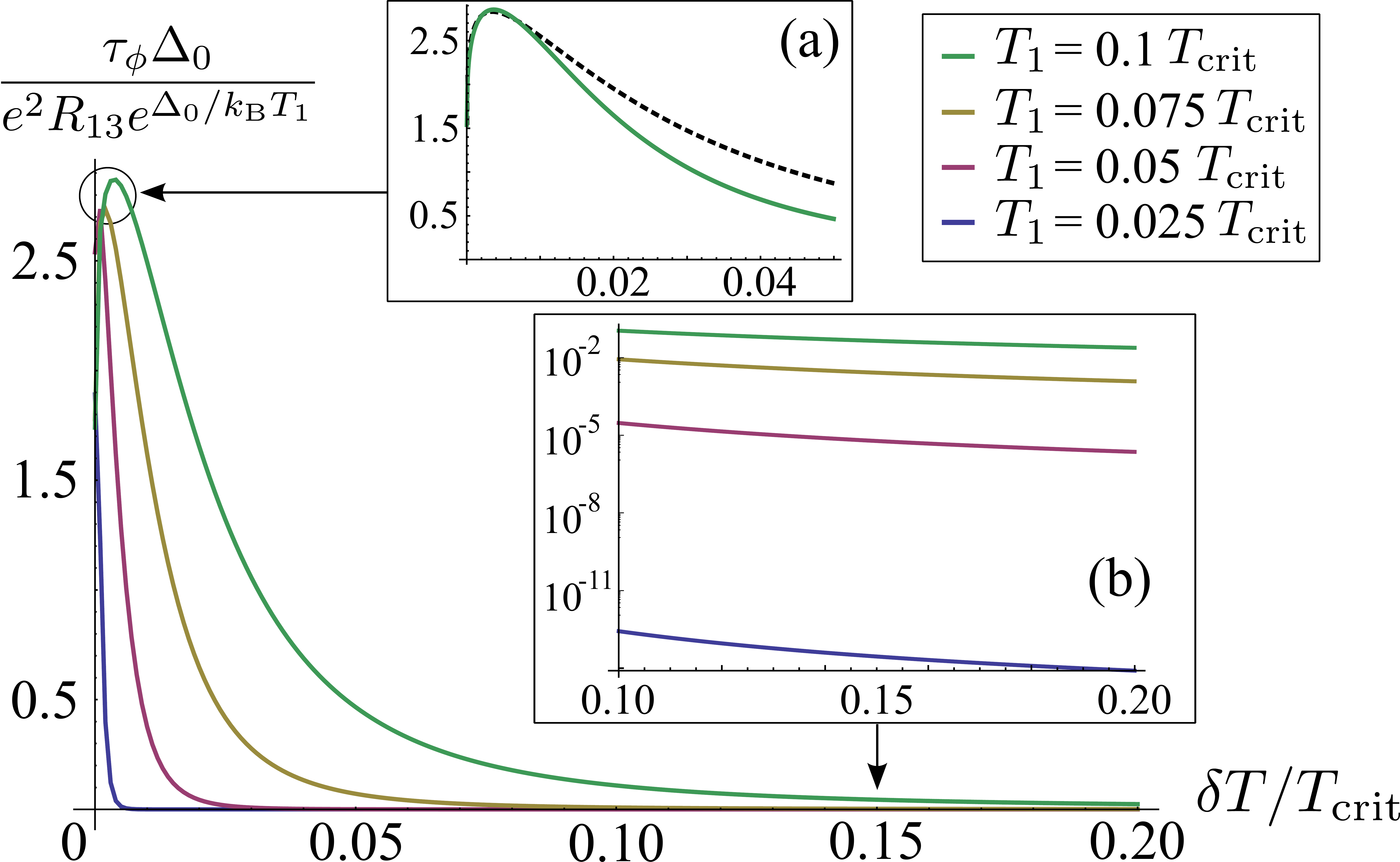}
\end{center}
\caption{Dephasing time as a function of the temperature gradient $\delta T$ for different values of $T_1$ with $T_1\leq T_3$. The dephasing time was calculated for $\alpha=0.75$ and $f=0.495$. In the inset (a) we show an enlargement at $\delta T\ll T$, for the case of $T_1=0.1T_\mathrm{crit}$ (full, green line) together with the approximate function of the dephasing time given in \eref{dephtimedT}  (dashed, black line) multiplied by a numerical factor of order 1. In the inset (b)  we show the enlargement of the plot for $0.1<\delta T/T_\mathrm{crit}<0.2$ on a logarithmic scale. Note that this plot is valid only for temperature differences larger than the cutoff temperature, $\delta T>T_\mathrm{cut}$, which in turn depends on the microscopic details of the Josephson junctions.}
\label{fig_DephTime}
\end{figure}
The dephasing time in units of $e^2R_{13} e^{\Delta_\mathrm{max}/k_\mathrm{B}T_1}/\Delta_0$ is shown in figure~\ref{fig_DephTime} as a function of the temperature difference $\delta T$ for different values of the minimum temperature $T_1$. The inset (a) of figure~\ref{fig_DephTime} compares the full result (full green line) with the approximation of equation~(\ref{dephtimedT}).

In the opposite regime of large temperature bias, we have that $r \simeq 1$
and thus the dephasing rate is approximately equal to $\gamma \simeq |\dot Q^3_\mathrm{L} - \dot
Q^3_\mathrm{R} |/ |\Delta_\mathrm{max}| $.  In particular, we find in this
regime that the dephasing time $\tau_\phi$ of the qubit is given by
\begin{equation}
  \tau_\phi 
  \simeq \frac{\Delta_0}{ | \dot Q^3_\mathrm{L} - \dot Q^3_\mathrm{R}
  |} \simeq \frac{ e^2 R_{13} e^{\Delta_0 /k_\mathrm{B}T_\mathrm{max}}}
  { \Delta_0 [C_\mathrm{L}-C_\mathrm{R}+\alpha(D_\mathrm{L}-D_\mathrm{R})]},
\end{equation} 
i.e.  the time after which the difference of the energy
transported by the heat currents in the two qubit states equals the gap
$\Delta_0$ of the superconductor. This is well confirmed by the inset (b) of figure~\ref{fig_DephTime}, which shows the dramatic decrease of the dephasing time with increasing temperature gradient, which in this regime is approximately given by $T_\mathrm{max}$.

Using the values we applied to estimate the heat currents in this system, we can
also estimate the dephasing time.  Taking $\Delta_0/e^2R_{13}\simeq1$THz as in
\cite{Giazotto12a}, we have that $\tau_\phi\approx1$ns for large temperature
gradients, $\delta T/T_\mathrm{min}\gg1$.  For small temperature gradients, it
can be shown that a temperature $T_\mathrm{min}$ of less than $0.1T_\mathrm{crit}$ has to be
reached in order to avoid a strong limitation of the dephasing due to the
thermal current.  Indeed, for $\delta T/T_\mathrm{min}\ll1$,
$T_\mathrm{min} \simeq T_\mathrm{max} \lesssim 0.1 T_\mathrm{crit}$, and taking the logarithm to be of
order one, we have $\tau_\phi\approx1\mu$s.  The actual dephasing times of the
Delft qubit range from a few tens of nanoseconds \cite{chiorescu} up to a
microsecond \cite{bertet} and thus are of the same order of magnitude.  As the
nominal temperatures reached for today's superconducting persistent current
qubits is usually smaller than $0.1 T_\mathrm{crit}$~\cite{Kemp11,Harris10},
it is unlikely that the thermal currents do  constitute the \textit{dominant} source
of dephasing for those qubits.  However, it is well known that quasiparticles
in small superconducting structures badly thermalise, leading to problems in
reaching the base temperature in the dilution
refrigerator~\cite{Riste13,Nguyen13} and thus effects of the phase dependence
of the thermal current on the coherence properties of the Delft qubit cannot be excluded.

\section{Conclusions}

We have shown that due to the phase sensitivity of the heat current which
flows in weak links of a superconducting loop, the heat current due to a temperature gradient applied to a flux qubit depends on
the state of the qubit which is formed when the loop is threaded with a
magnetic flux that is close to half a superconducting flux quantum.  We have
found that the sensitivity of the heat current to the qubit state can be up to
 $4\%$, when the qubit  is tuned away from the ``sweet spot''
of exactly half a flux quantum threading the loop. This should allow to identify the state of the flux qubit in experiments of the type performed in~\cite{Giazotto12,Giazotto12a}. 

Moreover, we have found
that due to this difference of heat currents at different qubit states, a thermal gradient leads to a dephasing of the qubit.  In
particular, we have found that the ratio of the dephasing rate to the difference of
the heat currents is universal with respect to microscopic details and only depends on the temperature of the
reservoirs measured in units of the superconducting gap at zero temperature.
For example, in the case of large temperature gradients the dephasing time of
the qubit corresponds to the time when the difference of heat currents have
transported an energy of the order of the superconducting gap.  
We have shown that the dephasing time of the flux qubit in the Delft design due
to the phase-sensitive heat current can range from microseconds for small
temperature differences to nanoseconds for large temperature differences thus
constituting a potential source of dephasing given the fact that the qubits
are driven by microwave pulses which may lead to an imbalance of heating
between the different sections of the superconducting loop.

\ack{We thank Gianluigi Catelani, Simone Gasparinetti, Francesco Giazotto, Dmitriy Golubev, and Anna Napoli for useful comments on our work. We would like to acknowledge financial support from the  Excellence Initiative of the German Federal and State Governments and the Ministry of Innovation NRW (S.S. and J.S.) and from the Alexander von Humboldt foundation (F.H.).}

\appendix

\section{Derivation of the Maki-Griffin formula for the heat current}
In this appendix, we derive in detail the analytic formulae for the heat current, which we use in (\ref{eq_Qqp}) and (\ref{eq_Qint}) and which were previously found in \cite{Maki65}.
The aim of this section is to describe the heat current in a superconductor-insulator-superconductor (SIS) Josephson junction, biased with a temperature 
gradient $\delta T$ across it, while no additional voltage is applied so that we have $\mu_1 = \mu_2 =\mu_0$. We assume that each superconductor (the two electrodes are denoted by $l=1,2$) is a 
particle reservoir in equilibrium at temperature $T_l$ and that it is characterised by the mean-field BCS Hamiltonian
\begin{equation}\label{SHam}
H_l=\sum_{k, \sigma} (\varepsilon_{l,k} -\mu_0)c^{ \dagger}_{l,k\sigma}c_{l,k\sigma}-\sum_{k}(\Delta_{l,k}c^{ \dagger}_{l,k\uparrow}c^{ \dagger}_{l,-k\downarrow}+\Delta^\ast_{l,k}c_{l,-k\downarrow}
c_{l,k\uparrow})\ .
\end{equation}
In (\ref{SHam}), $c^{ \dagger}_{l,k\sigma}$ and $c_{l,k\sigma}$ are single-electron creation and 
annihilation operators in the momentum $k$ and spin $\sigma$ representation and $\Delta_{l,k}$ is the superconducting energy gap of the  $l$-th electrode. Tunnelling between reservoirs is described by the tunnelling Hamiltonian
\begin{equation}\label{TunHam}
H_\mathrm{T}=\sum_{k,q,\sigma}(V^{12}_{kq}c^{ \dagger}_{1,k\sigma}c_{2,q\sigma}+V^{12\ast}_{kq}c^{ \dagger}_{2,q\sigma}c_{1,k\sigma}),
\end{equation}
where $k$ and $q$ are the momentum quantum numbers and the tunnelling matrix element is denoted by $V^{12}_{kq}$. The total Hamiltonian is then written as $H_\mathrm{tot}=H_1+H_2+H_\mathrm{T}$. 

Assuming that the system is sufficiently isolated and that in particular
phonons are frozen out at very low temperatures, the heat current in
electrode 1 is carried by electrons entering or leaving it, 
accompanied by a change in the overall energy $H_1$, with respect to the electrochemical potential.  According to the quantum-mechanical equation of motion, the heat current into the first electrode is
\begin{equation}\label{dQdt1}
\frac{dQ^{(1)}}{dt}=\left\langle\frac{d}{dt} H_1\right\rangle=\frac{i}{\hbar}\langle[H_\mathrm{tot}, H_1]\rangle.
\end{equation}
Of the full commutator $[H_\mathrm{tot}, H_1]$ only the contribution $[H_\mathrm{tot}, H_1]=[H_T, H_1 ]$ is non-zero.  Since we are dealing with fermionic annihilation and creation operators, they must obey the anticommutation rule $\{c^{ \dagger}_{l,k\sigma},c_{mk'\sigma'}
\}=\delta_{lm}\delta_{kk'}\delta_{\sigma\sigma'}$ from which we can easily derive
\begin{eqnarray}\label{CComm}
[c^{ \dagger}_{l,k\sigma}c_{l,k\sigma},c^{ \dagger}_{l,k'\sigma'}]
&=\delta_{kk'}\delta_{\sigma\sigma'}c^{ \dagger}_{l,k\sigma}, \quad
\big[c^{ \dagger}_{l,k \sigma}c_{l,k \sigma},c_{l,k' \sigma'} \big]
&=-\delta_{kk'}\delta_{\sigma\sigma'}c_{l,k\sigma}.
\end{eqnarray}
Using these commutation relations, the evaluation of $[H_T,H_1]$ yields
\begin{eqnarray}\label{Commutator}
  \fl \qquad
[H_T,H_1]
&=\sum_{k,q, \sigma}\sum_{k', \sigma'}\Big\{ V^{12}_{kq} \xi_{1,k'} [c^{ \dagger}_{1,k\sigma},c^{ \dagger}_{1,k'\sigma'}c_{1,k'\sigma'}]c_{2,q\sigma}+V^{12\ast}_{kq} \xi_{1,k'} c^{\dagger}_{2,q\sigma}[c_{1,k\sigma},c^{ \dagger}_{1,k'\sigma'}c_{1,k'\sigma'}]\nonumber\\
&\quad-V^{12}_{kq}\Delta^*_{1,k'}[c^{ \dagger}_{1,k\sigma},c_{1,-k'\downarrow}c_{1,k'\uparrow}]c_{2,q\sigma}-V^{12\ast}_{kq}\Delta_{1,k'}c^{\dagger}_{2,q\sigma}[c_{1,k\sigma},c^{ \dagger}_{1,k'\uparrow}c^{ \dagger}_{1,-k'\downarrow}]\Big\}\nonumber\\
&=-2 i \,\mathrm{Im}\Big\{\sum_{k,q, \sigma}V^{12}_{kq} \xi_{1,k} c^{ \dagger}_{1,k\sigma}c_{2,q\sigma}+V^{12}_{kq}(\Delta^*_{1,-k}c_{1,-k\uparrow}c_{2,q\downarrow}-\Delta^*_{1,k}c_{1,-k\downarrow}c_{2,q\uparrow})\Big\}.\nonumber\\
\end{eqnarray}
Substituting this expression in (\ref{dQdt1}), the heat current is found to be
\begin{equation}\label{dQdt2}
\fl\frac{dQ^{(1)}}{dt}=\frac{2}{\hbar}\,\mathrm{Im}\left\{\sum_{k,q, \sigma}\Big\langle V^{12}_{kq} \xi_{1,k} c^{ \dagger}_{1,k\sigma}c_{2,q\sigma}+V^{12}_{kq}(\Delta^*_{1,-k}c_{1,-k\uparrow}c_{2,q\downarrow}-\Delta^*_{1,k}c_{1,-k\downarrow}c_{2,q\uparrow})\Big\rangle\right\}.
\end{equation}
The next step is to calculate the expectation values in the general expression for the heat current (\ref{dQdt2}), yielding a Kubo formula, when expanding in the  small tunnelling matrix elements. In general, to first order in perturbation theory, the expectation value of an operator $O(t)$ is
\begin{equation}\label{Kubo}
\langle O(t)\rangle=-i\int_{-\infty}^t dt'\langle [O(t),H_T(t')]\rangle_0 e^{\eta (t'-t)}
\end{equation}
where the brackets $\langle \cdot\rangle_0$ denote the equilibrium average with respect to the Hamiltonian $H_0=H_1+H_2$ without the perturbation $H_\mathrm{T}$, and $\eta$ is a small  parameter which is eventually taken to zero. Using (\ref{Kubo}), the heat current can be written as 
\begin{eqnarray}\label{dQdt2.1}
  \fl \qquad \qquad
\frac{dQ^{(1)}}{dt}&=-\frac{2}{\hbar}\,\mathrm{Re}\bigg\{\int_{-\infty}^tdt'e^{\eta( t'-t)}\sum_{k,q, \sigma}\Big\langle\Big[\big(V^{12}_{kq} \xi_{1,k}c^{ \dagger}_{1,k\sigma}(t)c_{2,q\sigma}(t)+	\nonumber\\
&\quad+V^{12}_{kq}(\Delta^*_{1,-k}c_{1,-k\uparrow}(t)c_{2,q\downarrow}(t)-\Delta^*_{1,k}c_{1,-k\downarrow}(t)c_{2,q\uparrow}(t))\big),H_T(t')\Big]\Big\rangle_0\bigg\}.
\end{eqnarray}
As a first step, we need to again evaluate the commutator expression in the integrand, which assumes the form
\begin{eqnarray}\label{Comm2}
\fl\sum_{k,q,\sigma}\sum_{k',q',\sigma'}\bigg\{V^{12}_{kq}V^{12}_{k'q'}\Big[\xi_{1,k}\Big(c^\dagger_{1,k\sigma}(t)c_{2,q\sigma}(t)c^\dagger_{1,k'\sigma'}(t')c_{2,q'\sigma'}(t')-c^\dagger_{1,k'\sigma'}(t')c_{2,q'\sigma'}(t')c^\dagger_{1,k\sigma}(t)c_{2,q\sigma}(t)\Big)\nonumber\\
\fl\quad-\Delta_{1,k}^*\Big(c_{1,-k\downarrow}(t)c_{2,q\uparrow}(t)c^\dagger_{1,k'\sigma'}(t')c_{2,q'\sigma'}(t')-c^\dagger_{1,k'\sigma'}(t')c_{2,q'\sigma'}(t')c_{1,-k\downarrow}(t)c_{2,q\uparrow}(t)\Big)\nonumber\\
\fl\quad+\Delta_{1,-k}^*\Big(c_{1,-k\uparrow}(t)c_{2,q\downarrow}(t)c^\dagger_{1,k'\sigma'}(t')c_{2,q'\sigma'}(t')-c^\dagger_{1,k'\sigma'}(t')c_{2,q'\sigma'}(t')c_{1,-k\uparrow}(t)c_{2,q\downarrow}(t)\Big)\Big]\nonumber\\
\fl\quad+V^{12}_{kq}V^{12\ast}_{k'q'}\Big[\xi_{1,k}\Big(c^\dagger_{1,k\sigma}(t)c_{2,q\sigma}(t)c^\dagger_{2,q'\sigma'}(t')c_{1,k'\sigma'}(t')-c^\dagger_{2,q'\sigma'}(t')c_{1,k'\sigma'}(t')c^\dagger_{1,k\sigma}(t)c_{2,q\sigma}(t)\Big)\nonumber\\
\fl\quad-\Delta_{1,k}^*\Big(c_{1,-k\downarrow}(t)c_{2,q\uparrow}(t)c^\dagger_{2,q'\sigma'}(t')c_{1,k'\sigma'}(t')-c^\dagger_{2,q'\sigma'}(t')c_{1,k'\sigma'}(t')c_{1,-k\downarrow}(t)c_{2,q\uparrow}(t)\Big)\nonumber\\
\fl\quad+\Delta_{1,-k}^*\Big(c_{1,-k\uparrow}(t)c_{2,q\downarrow}(t)c^\dagger_{2,q'\sigma'}(t')c_{1,k'\sigma'}(t')-c^\dagger_{2,q'\sigma'}(t')c_{1,k'\sigma'}(t')c_{1,-k\uparrow}(t)c_{2,q\downarrow}(t)\Big)\Big]\bigg\}.
\end{eqnarray}
In order to take the equilibrium expectation value of this expression,
it is useful to employ the Green's functions defined in the following way
\begin{eqnarray*}
G^>_{l,k}(t,t')&=-i\langle c_{l,k}(t)c^\dagger_{l,k}(t')\rangle_0,
&G^<_{l,k}(t,t')=i\langle c^{\dagger}_{l,k}(t')c_{l,k}(t)\rangle_0,\nonumber\\
F^>_{l,k}(t,t')&=-i\langle c_{l,k\uparrow}(t)c_{l,-k\downarrow}(t')\rangle_0,
\quad
&F^<_{l,k}(t,t')=i\langle c_{l,-k\downarrow}(t')c_{l,k\uparrow}(t)\rangle_0,\nonumber\\
F^{\dagger >}_{l,k}(t,t')&=i\langle c^{\dagger}_{l,-k\downarrow}(t')c^{\dagger}_{l,k\uparrow}(t)\rangle_0,
&F^{\dagger <}_{l,k}(t,t')=-i\langle c^{\dagger}_{l,k\uparrow}(t)c^{\dagger}_{l,-k\downarrow}(t')\rangle_0.\nonumber
\end{eqnarray*}
The linear-response formula for the heat current, equation (\ref{dQdt2.1}) is then given by
\begin{eqnarray}\label{dQdt3}
\fl \frac{dQ^{(1)}}{dt}=\frac{2}{\hbar}\,\mathrm{Re}\sum_{k,q}\int_{-\infty}^tdt'e^{\eta (t'-t)}\bigg\{V^{12}_{kq}V^{12}_{-k-q}\Big[\xi_{1,k}\Big(F^{\dagger >}_{1,k}(t,t')F^{<}_{2,q}(t,t')+F^{\dagger <}_{1,-k}(t',t)F^{>}_{2,-q}(t',t)\nonumber\\
\fl\quad-F^{\dagger <}_{1,k}(t,t')F^{>}_{2,q}(t,t')-F^{\dagger >}_{1,-k}(t',t)F^{<}_{2,-q}(t',t)\Big)+\Delta^*_{1,k}\Big(G^>_{1,-k}(t,t')F^{>}_{2,q}(t,t')\nonumber\\
\fl\quad-G^<_{1,-k}(t,t')F^{<}_{2,q}(t,t')\Big)+\Delta^*_{1,-k}\Big(G^>_{1,-k}(t,t')F^{<}_{2,-q}(t',t)-G^<_{1,-k}(t,t')F^{>}_{2,-q}(t',t)\Big)\Big]\nonumber\\
\fl\quad-|V^{12}_{kq}|^2\Big[2\xi_{1,k}\Big(G^<_{1,k}(t',t)G^>_{2,q}(t,t')-G^>_{1,k}(t',t)G_{2,q}^<(t,t')\Big)+\Delta_{1,k}^*\Big(F^<_{1,k}(t',t)G^>_{2,q}(t,t')\nonumber\\
\fl\quad-F^>_{1,k}(t',t)G^<_{2,q}(t,t')\Big)+\Delta_{1,-k}^*\Big(F^>_{1,-k}(t,t')G^>_{2,q}(t,t')-F^<_{1,-k}(t,t')G^<_{2,q}(t,t')\Big)\Big]\bigg\}.
\end{eqnarray}
The next step is to express the Green's functions by their spectral densities
\begin{eqnarray}\label{GreenF}
G^>_{l,k}(t,t')&=-i\int_{-\infty}^{\infty}\frac{d\omega}{2\pi}e^{-i\omega(t-t')}(1-f_{l}(\omega))A_{l,k}(\omega),\nonumber\\
G^<_{l,k}(t,t')&=i\int_{-\infty}^{\infty}\frac{d\omega}{2\pi}e^{-i\omega(t-t')}f_{l}(\omega)A_{l,k}(\omega),\nonumber\\
F^>_{l,k}(t,t')&=i\int_{-\infty}^{\infty}\frac{d\omega}{2\pi}e^{-i\omega(t-t')}(1-f_{l}(\omega))B_{l,k}(\omega),\nonumber\\
F^<_{l,k}(t,t')&=-i\int_{-\infty}^{\infty}\frac{d\omega}{2\pi}e^{-i\omega(t-t')}f_{l}(\omega)B_{l,k}(\omega).
\end{eqnarray}
where $f_{l}(\omega)$ is the Fermi function of the $l$-th electrode.  Substituting these expressions, (\ref{GreenF}), into the equation for the heat current, (\ref{dQdt3}), the latter simplifies significantly
\begin{eqnarray}\label{dQdt4}
\fl\frac{dQ^{(1)}}{dt}&=\frac{2}{\hbar}\,\mathrm{Im}\sum_{k,q}\int_{-\infty}^{\infty}\frac{d\omega}{2\pi}\int_{-\infty}^{\infty}\frac{d\omega'}{2\pi}|V^{12}_{kq}|^2\big(f_{1}(\omega)-f_{2}(\omega')\big)\nonumber\\
\fl&\quad\times\Bigg(-\Delta^*_{1,-k}\frac{A_{1,-k}(\omega)B_{2,-q}(\omega')}{\omega'-\omega-i\eta}+\Delta^*_{1,k}\frac{A_{1,k}(\omega)B_{2,q}(-\omega')}{\omega'-\omega+i\eta}+2\Delta^*_{1,k}\frac{B_{1,k}(\omega)A_{2,q}(\omega')}{\omega'-\omega+i\eta}\nonumber\\
\fl&\quad+\xi_{1,k}\frac{B^*_{1,k}(\omega)B_{2,q}(\omega')-B_{1,k}(\omega)B^*_{2,q}(\omega')}{\omega'-\omega-i\eta}+2\xi_{1,k}\frac{A_{1,k}(\omega)A_{2,q}(\omega')}{\omega'-\omega-i\eta}\Bigg)\ .
\end{eqnarray}
Here, since the tunneling matrix element is invariant under time reversal, we used the relation $V^{12}_{kq}V^{12}_{-k-q}=|V^{12}_{kq}|^2$.
According to microscopic BCS theory the spectral densities are $A_{l,k}(\omega)=2\pi[|u_{l,k}|^2\delta(\omega-E_{l,k})+|v_{l,k}|^2\delta(\omega+E_{l,k})]$ and $B_{l,k}(\omega)=2\pi u_{l,k}v_{l,k}[\delta(\omega-E_{l,k})-\delta(\omega+E_{l,k})]$, with $|u_{l,k}|^2=1/2(1+\xi_{l,k}/E_{l,k})$, $|v_{l,k}|^2=1/2(1-\xi_{l,k}/E_{l,k})$, and the quasi-particle energy-momentum relation
$E_{l,k}=\sqrt{\xi_{l,k}^2+|\Delta_{l,k}|^2}$.
To continue the calculation, it is important to notice that the parameters
$u_{l,k}$, $v_{l,k}$ and $\Delta_{l,k}$ are not independent, but that their phases are related by $\Delta^*_{l,k}v_{l,k}/u_{l,k}=E_{l,k}-\xi_{l,k}$, such that
 $\Delta^*_{l,k}v_{l,k}/u_{l,k}$ must be a real number. That is, the phase of $v_{l,k}$ relative to $u_{l,k}$ must be equal to the phase of $\Delta_{l,k}$. Without  loss of generality we can choose $u_{l,k}$ to be real and positive, so that $v_{l,k}$ and $\Delta_{l,k}$ must have the same phase~\cite{Tinkham}.
Finally, we introduce the phase difference $\varphi$ between the electrodes with the relation
$\Delta^*_{1,k}v_{2,q}=|\Delta^*_{1,k}v_{2,q}|\exp{(i\varphi)}$.
 
 The next stage of the calculation is to substitute the spectral densities, $A_{l,k}(\omega)$ and $B_{l,k}(\omega)$, into the heat current expression,  
(\ref{dQdt4}), and to perform the sum over the momenta $k$ and $q$. To do that, the sum over the momenta is transformed into an integral over the electronic energies $\xi_{l,k}$ with $l=1,2$, such that~\cite{Bruus},
\begin{eqnarray}\label{SpecDens2}
\sum_{k}A_{l,k}&=\frac{\pi N^0_l}{2}\int_{-\infty}^{\infty}d\xi_{l,k}\bigg(\delta(\omega-E_{l,k})+\delta(\omega+E_{l,k})\bigg)\\
&=\pi
N^0_l\int_{|\Delta_{l,k}|}^{\infty}dE_{l,k}\frac{E_{l,k}}{\sqrt{E_{l,k}^2-|\Delta_{l,k}|^2}}\bigg(\delta(\omega-E_{l,k})+\delta(\omega+E_{l,k})\bigg)\nonumber\\
&=\frac{\pi
N^0_l|\omega|}{\sqrt{\omega^2-|\Delta_{l}|^2}}\theta(\omega^2-|\Delta_{l}|^2)\nonumber \ .
\end{eqnarray}
We denote the normal-state density of states (including spin) of the
$l$-th electrode by $N^0_{l}$ and finally assumed an  isotropic superconductor with an energy-independent gap.
Similarly, we find  for the other terms
\begin{eqnarray}\label{SpecDens3}
\sum_kB_{l,k}&=\mathrm{sgn}(\omega)\frac{\pi
N^0_l|\Delta_{l}|}{\sqrt{\omega^2-|\Delta_{l}|^2}}\theta(\omega^2-|\Delta_{l}|^2),\nonumber\\
\sum_k\xi_{l,k} A_{l,k}&=\mathrm{sgn}(\omega)\frac{\pi
N^0_l(\omega^2-|\Delta_{l}|^2)}{\sqrt{\omega^2-|\Delta_{l}|^2}}\theta(\omega^2-|\Delta_{l}^2),\\
\sum_k\xi_{l,k} B_{l,k}&=0.\nonumber
\end{eqnarray}
In (\ref{dQdt4}) the term depending only on $B_{1,k}B_{2,q}$, which is related to the sole Cooper pairs contribution,  vanishes, as it easy to notice using the last expression in (\ref{SpecDens3}).
If we also use the relation $\lim_{\epsilon \rightarrow0}\mathrm{Im}\left\{1/(x-i\epsilon)\right\}=\pi\delta(x)$
and substitute the integrals over the spectral functions, (\ref{SpecDens2}) and (\ref{SpecDens3}), into the heat current, 
(\ref{dQdt4}), we obtain
\begin{eqnarray}\label{dQdt5}
\frac{dQ^{(1)}}{dt}&=\frac{\pi}{\hbar}|V^{12}_{kq}|^2N^0_1N^0_2\int_{-\infty}^{\infty}\!\!\!\!
d\omega\,\omega\frac{\big(f_{1}(\omega)-f_{2}(\omega)\big)\theta(\omega^2-|\Delta_{1}|^2)\theta(\omega^2-|\Delta_{2}|^2)}{\sqrt{\omega^2-|\Delta_{1}|^2}\sqrt{\omega^2-|\Delta_{2}|^2}}\nonumber\\
&\qquad\times
\biggl[\omega^2-|\Delta_{1}||\Delta_{2}|\cos\varphi\biggr].
\end{eqnarray}
In the above derivation we assumed that the normal densities of states and the tunnelling matrix
elements are energy-independent. Evaluating the theta-functions we finally obtain
\begin{eqnarray}\label{dQdt6}
  \fl
\frac{dQ^{(1)}(T_1,T_2)}{dt}&=\frac{d \Qqp^{(1)}(T_1,T_2)}{dt}-\frac{d \Qint^{(1)}(T_1,T_2)}{dt}\cos{\varphi}\\
\fl
&=\frac{2}{e^2R_{12}}\int_{|\Delta_\mathrm{max}|}^{\infty}d\omega\omega\frac{\big(f_1(\omega)-f_2(\omega)\big)}{\sqrt{\omega^2-|\Delta_{1}|^2}\sqrt{\omega^2-|\Delta_{2}|^2}}\bigg[\omega^2-|\Delta_{1}||\Delta_{2}|\cos\varphi\bigg]\nonumber
\end{eqnarray}
where we introduced $|\Delta_\mathrm{max}|=\mathrm{max}\{|\Delta_1(T_1)|,|\Delta_2(T_2)|\}$ and
the normal-state conductance of the Josephson junction, defined via the inverse of the normal-state resistance, $R_{12}^{-1}=\pi e^2 N^0_1 N^0_2 |V^{12}_{kq}|^2/\hbar$.

In ({\ref{dQdt6}), the total heat current through the junction carried by quasiparticles is $dQ_\mathrm{qp}^{(1)}(T_1,T_2)/dt$, while $dQ_\mathrm{int}^{(1)}(T_1,T_2)/dt$ is  the interference contribution to the heat current due to an interplay between quasiparticles and
Cooper pair condensate. It is easy to see that $d\Qint^{(1)}(T_1,T_2)/dt$, which originates from the Josephson effect and is characteristic to weakly coupled superconductors, vanishes when at least one of the superconductors is in the normal state ($|\Delta_l(T_l)|=0$).

\section{Slopes of the sensitivities}\label{app_slopes}

In this section of the appendix we provide the analytic formulas for the slopes of the sensitivities $s_l$, for small $\delta f\ll1$. We find
\begin{eqnarray*}
  \fl
m_1&=\frac{2\pi\bigg[(4\alpha^2-1)^2(2\alpha^2-1)+\alpha \frac{E_\mathrm{C}}{E_\mathrm{J}}(1+2\alpha^2)\bigg]}{(4\alpha^2-1)^{5/2}\bigg[1-2\alpha\frac{\dot{Q}_\mathrm{qp}}{\dot{Q}_\mathrm{int}}e^{\frac{E_\mathrm{C}}{E_\mathrm{J}}\frac{\alpha }{(4\alpha^2-1)}}\bigg]}\\
\fl
m_2&=\frac{2\pi\alpha\bigg[(4\alpha^2-1)+4\alpha \frac{E_\mathrm{C}}{E_\mathrm{J}}\bigg]}{(4\alpha^2-1)^{3/2}\bigg[(2\alpha^2-1)-2\alpha^2\frac{\dot{Q}_\mathrm{qp}}{\dot{Q}_\mathrm{int}}e^{ \frac{E_\mathrm{C}}{E_\mathrm{J}}\frac{4\alpha}{(4\alpha^2-1)}}\bigg]}\\
\fl
m_3&=\frac{
                2\pi\alpha\left[
                                 (4\alpha^2-1)^2\bigg(2\alpha^2-1+e^{\frac{E_\mathrm{C}}{E_\mathrm{J}}\frac{3\alpha }{(4\alpha^2-1)}}\bigg)
                                 +\alpha \frac{E_\mathrm{C}}{E_\mathrm{J}}\bigg(2\alpha^2+1+e^{\frac{E_\mathrm{C}}{E_\mathrm{J}}\frac{4\alpha}{(4\alpha^2-1)}}(4\alpha^4-1)\bigg)\right]}
                                 {(4\alpha^2-1)^{5/2}\left[\bigg(e^{\frac{E_\mathrm{C}}{E_\mathrm{J}}\frac{3\alpha }{(4\alpha^2-1)}}(2\alpha^2-1)+1\bigg)-2\alpha(\alpha+1)\frac{\dot{Q}_\mathrm{qp}}{\dot{Q}_\mathrm{int}}e^{-\frac{E_\mathrm{C}}{E_\mathrm{J}}\frac{\alpha }{(4\alpha^2-1)}}\right]}.
\end{eqnarray*}
The results are plotted in figure~\ref{fig_m_alpha}.
\section*{Bibliography}

\bibliographystyle{iopart-num}

\providecommand{\newblock}{}

\end{document}